\documentstyle[11pt,newpasp,twoside,epsf]{article}
\def\edcomment#1{\iffalse\marginpar{\raggedright\sl#1\/}\else\relax\fi}
\reversemarginpar

\begin{document}
\title{Dual Axisymmetry in  
Proto-Planetary Nebula Reflection Nebulosities:
Results from an HST Snapshot Survey of PPN
Candidates}
\author{Toshiya Ueta and Margaret Meixner}
\affil{Department of Astronomy, MC-221, 
University of Illinois at Urbana-Champaign, 
Urbana, IL  61801}
\author{Matthew Bobrowsky}
\affil{Challenger Center for Space Science Education,
1250 North Pitt Street,
Alexandria, VA  22314}

\begin{abstract}
We summarize results of our {\it HST} imaging survey of 
proto-planetary nebula (PPN) candidates, in which we discovered 
two types of axisymmetric reflection nebulosities.
The Star-Obvious Low-level-Elongated (SOLE) nebulae show 
a bright central star embedded in a smooth, faint extended 
nebulosity, whereas the DUst-Prominent Longitudinally-EXtended
(DUPLEX) nebulae have remarkable bipolar structure with 
a completely or partially obscured central star.
The intrinsic axisymmetry of the PPN reflection nebulosities 
demonstrates that the axisymmetry often
found in planetary nebulae predates the PPN phase.
We suggest that the major cause for the apparent dual 
morphology is the optical depth difference in the circumstellar
dust shell rather than the inclination angle effect.
\end{abstract}

\section{Introduction}
Axisymmetry often seen in planetary nebulae (PNe) must arise
before photoionization starts illuminating the fascinating 
structure of the nebulae.
High resolution optical imaging of PPN reflection nebulosities 
can serve as an indirect probe for the innermost structure of 
the PPN dust shell, which is formed during the superwind mass 
loss phase at the end of the asymptotic giant branch (AGB) 
evolution.
This conference contribution summarizes the results from
our {\it HST} survey, 
which covered the largest number of PPN candidates to date
including the ones associated with bright central stars
(Ueta, Meixner, \& Bobrowsky 2000).
Our goal was to investigate if there exists any coherent
trend that will bridge gaps in the circumstellar 
morphologies between the AGB and PN phases.

\section{HST Snapshot Survey of PPN Candidates}
Our HST snapshot survey of PPN candidates found that 78\% 
(21 of 27) of the reflection nebulosities were resolved and 
all of the resolved nebulae were elongated (ellipticity $=0.44$).
This ubiquitous axisymmetry suggests that PPNe are intrinsically 
axisymmetric.
Hence, given the spherical nature of the AGB wind shell, 
the PN axisymmetry is likely to originate during the superwind 
phase.
Moreover, there are clearly two types of reflection nebulosities 
which we classified as 
{\it Star-Obvious Low-level-Elongated} (SOLE) nebulae,
showing smooth, low surface brightness
elongations with extremely bright central stars,
and 
{\it DUst-Prominent Longitudinally-EXtended} (DUPLEX) nebulae,
having bipolar lobes straddling dust lanes with
totally or partially obscured central stars.

\begin{figure}[h]
\plotone{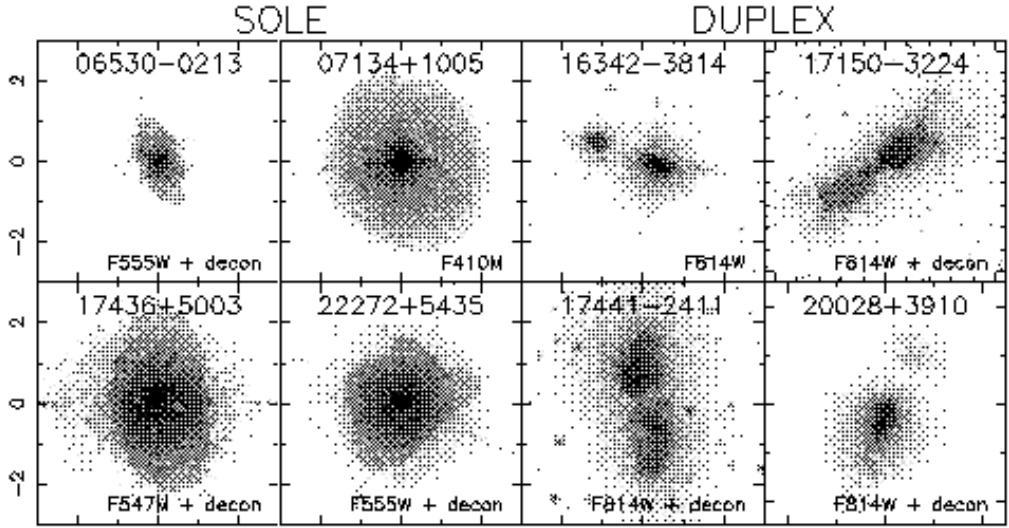}
\caption{SOLE (left four frames) and DUPLEX (right four frames) 
types of PPN reflection nebulae.
IRAS ID and filters are shown in each frame.
N (E) is up (left) with tickmarks showing relative offsets 
in arcsec.
``+ decon'' indicates a deconvolved image.}
\end{figure}

\section{SOLE vs. DUPLEX: Physically Distinct Nebulosities}

We attribute the {\it major} source for the dual morphology 
to the optical depth in the superwind shell, which varies
depending on the degree of equatorial density enhancement.
The following evidence suggests that the inclination 
angle effect alone may not be good enough to interpret what we 
see in the data.

\subsection{Mid-Infrared Morphology}
A recent mid-infrared (mid-IR) survey of PPN candidates 
(Meixner et al. 1999) 
has revealed a corresponding dual morphology in the PPN
dust emission regions.
SOLE type optical nebulae corresponds to {\it toroidal} type 
mid-IR dust emission nebulae, in which we see two dust emission
peaks that are interpreted as a limb-brightened dust torus.
On the other hand, DUPLEX type nebulae
is related to {\it core/elliptical} type dust emission nebulae, 
in which we see a very compact, unresolved core with a broad 
plateau (Fig.2).
Characteristics of both optical and mid-IR images
suggest that SOLE-Toroidal type nebulae are of
optically thin dust shells and
that DUPLEX-Core/Elliptical type nebulae are of
optically thick dust shells.
We also schematically describe how the optical depth
in the superwind influences the shape of the
two types of reflection nebulosities in Fig.2.

\begin{figure}[h]
\plottwo{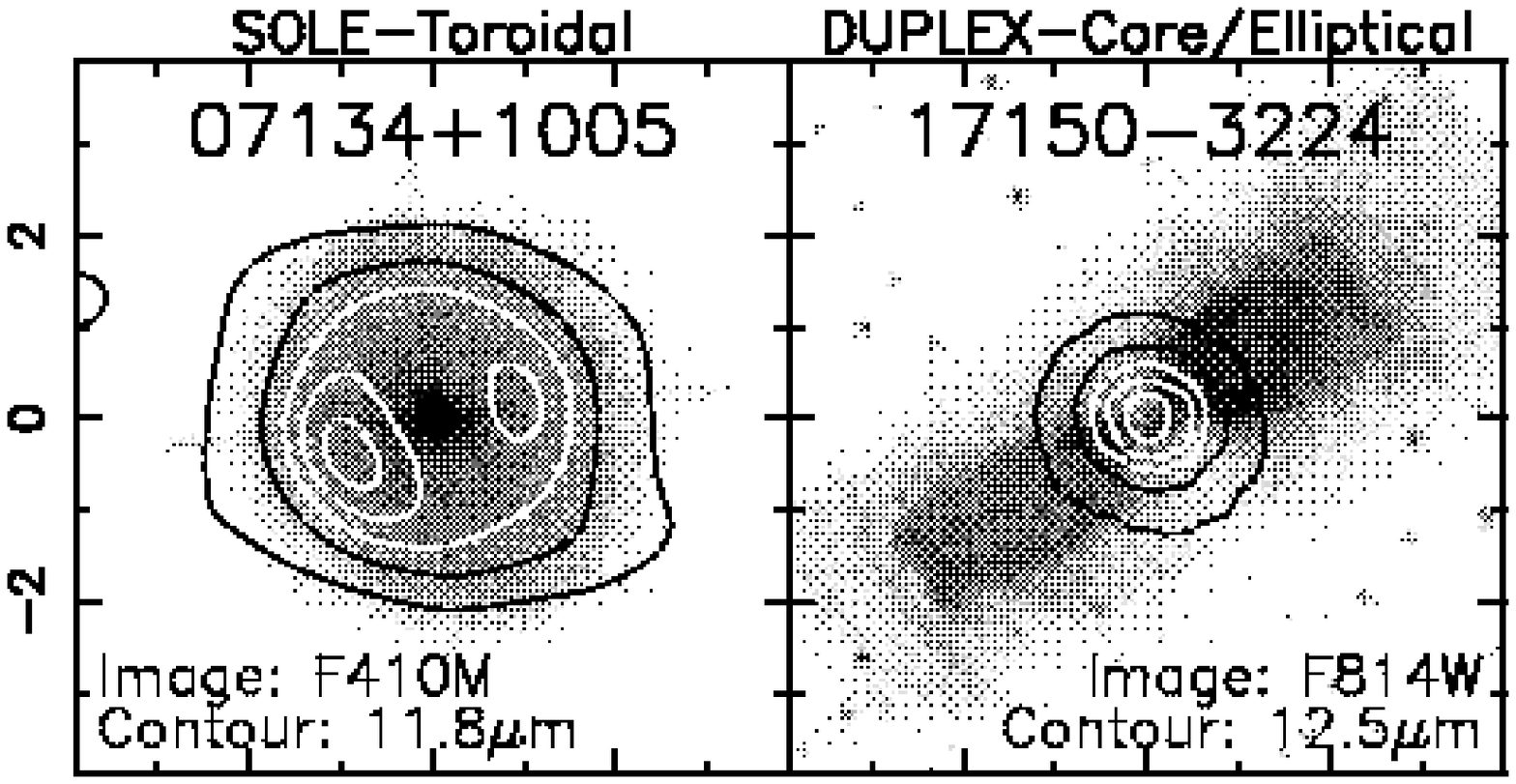}{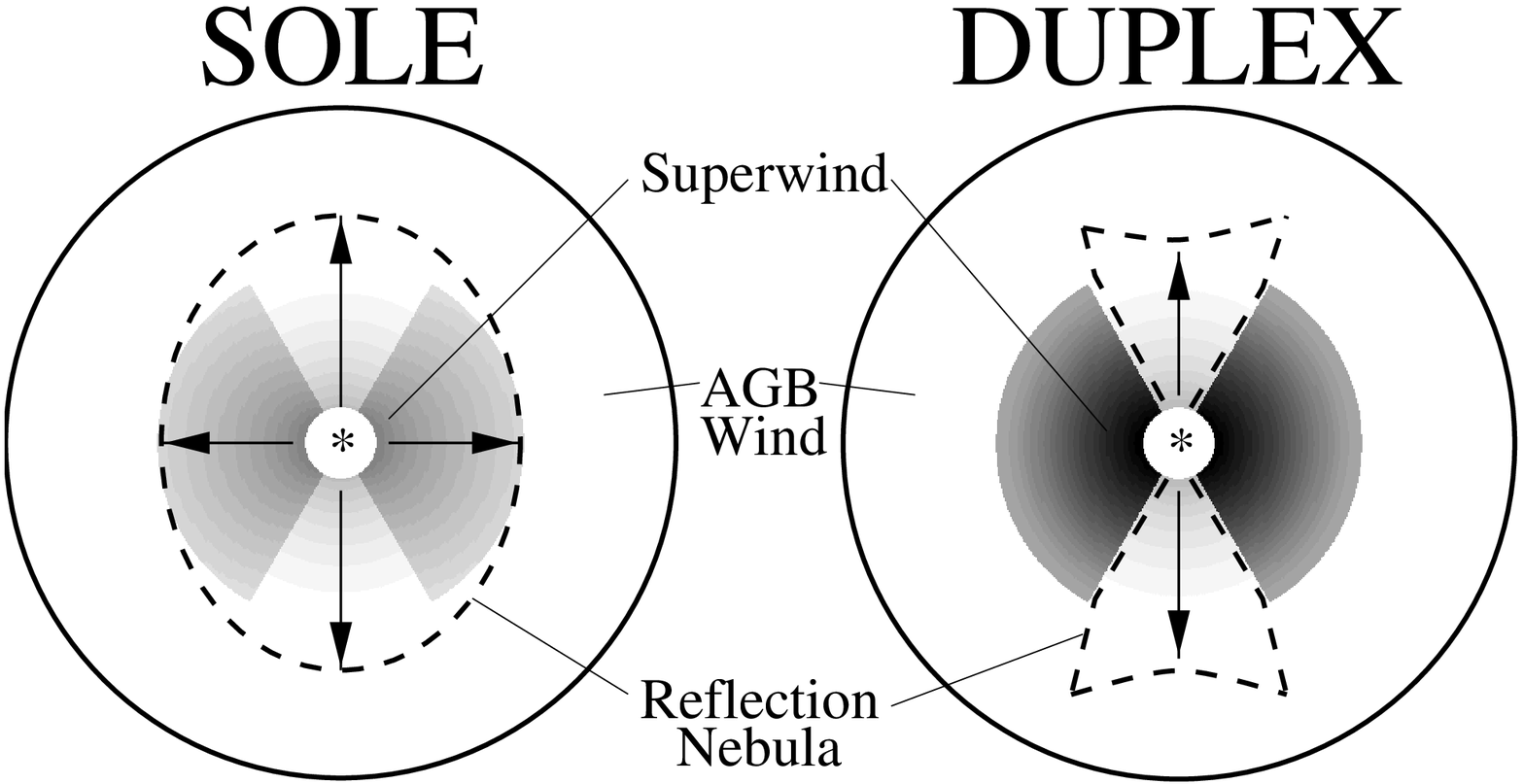}
\caption{Optical/mid-IR overlay images (left) and 
a schematic (right) of two PPN types.
Both optical and mid-IR PPN morphologies appear to
depend on the optical depth of the superwind shell.}
\end{figure}

More importantly, mid-IR images show that there are SOLE 
nebulae that are oriented rather edge-on (e.g. IRAS 07134+1005 
in Fig.2, IRAS 17436+5003 in Fig.4), and therefore, 
elliptical reflection nebulae (SOLE) are not necessarily 
bipolar nebulae (DUPLEX) seen pole-on.

\subsection{Spectral Energy Distributions and Two-Color Diagrams}
SOLE and DUPLEX nebulae are also distinct in the shapes of 
spectral energy distribution (SED; Fig.3, left).
Optically thin SOLE nebulae let the stellar 
emission pass while converting some to thermal dust emission,
yielding comparable stellar and dust peaks.
Optically thick DUPLEX nebulae
absorb all stellar photons except for some scattered 
ones, yielding a large dust emission peak with
an optical plateau.

\begin{figure}[b]
\plotone{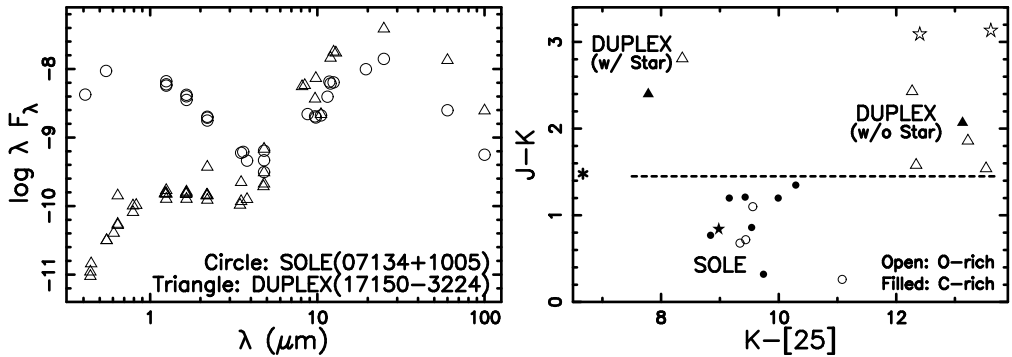}
\caption{The optical depth difference in SOLE and DUPLEX 
nebulae manifests itself as the SED shapes (left) and 
color diagram (right).}
\end{figure}

The duality is also seen in both an IRAS/near-IR 
and a near-IR color diagrams,
confirming the optical thickness interpretation.
IRAS/near-IR diagram (Fig.3, right) shows
three clusters of PPNe according to the visibility 
of the central stars (total, partial, and none obscuration).
Near-IR diagram (not shown) shows
a linear distribution of PPNe according to the degree 
of reddening (the redder, the more dust).

\subsection{Two-Dimensional Radiation Transfer Calculations}
Preliminary results from full 2-D radiation transfer 
simulations of PPN dust shells generally agree
with optical/mid-IR data. 
The optically thin superwind shell model yielded a resolved
dust peak in the mid-IR and a diffuse reflection nebula without 
a dust lane in the optical, 
while the optically thick superwind shell model
yielded a compact dust emission core with a broad emission
plateau in the mid-IR and a bipolar nebula divided by a dust lane
in the optical (Meixner, Ueta, \& Bobrowsky 2000;
also see the contribution by Meixner in this issue).

\section{Discussions}
The distinctness between SOLE and DUPLEX nebulae seems to
originate from the varying optical depths in the superwind shells,
and it is evidenced by the correlation between optical and mid-IR
morphologies, characteristic SED shapes,
color diagrams,
and 2-D radiation transfer calculations.
The inclination angle effect alone may not be good enough to 
interpret all SOLE nebulae 
being pole-on bipolar nebulae because
we have found as many SOLE nebulae as DUPLEX nebulae, 
there are some SOLE PPNe that are rather edge-on,
and SOLE nebulae do not show the imbalance of surface brightness, 
which is a signature of inclined orientation in DUPLEX nebulae,
although the inclination angle effect still remains as a source 
of confusion.

Neither age nor chemical composition seems to be related to the 
morphological bifurcation, and the origin of equatorial density 
enhancement in the superwind remains unknown.
However, there is a suggestion that DUPLEX PPNe may have 
evolved from higher mass AGB progenitors because of the Galactic 
distribution of PPNe:
DUPLEX nebulae, having a mean height of $220$pc with a range 
of $|z|<520$pc, are more confined to the Galactic plane
than SOLE nebulae, which have a mean height of
$470$pc with a range of $|z|<2100$pc.

\end{document}